\documentclass[12pt]{article}%
\usepackage{amssymb}%
\usepackage{amsmath}%
\usepackage{amsfonts}%
\usepackage{graphicx}
\providecommand{\U}[1]{\protect\rule{.1in}{.1in}}

\begin{document}

\title{Nickel on Lead, Magnetically Dead or Alive?}
\author{Go Tateishi and Gerd Bergmann\\Department of Physics\\University of Southern California\\Los Angeles, California 90089-0484\\e-mail: bergmann@usc.edu}
\date{\today}
\maketitle

\begin{abstract}
Two atomic layers of Ni condensed onto Pb films behave, according to anomalous
Hall effect measurements, as magnetic dead layers. However, the Ni lowers the
superconducting $T_{c}$ of the Pb film. This has lead to the conclusion that
the Ni layers are still very weakly magnetic. In the present paper the
electron dephasing due to the Ni has been measured by weak localization. The
dephasing is smaller by a factor 100 than the pair-breaking. This proves that
the $T_{c}$-reduction in the PbNi films is not due magnetic Ni moments.


\end{abstract}

When a thin superconductor is condensed onto a normal conducting film then the
first layers loose their superconductivity. This phenomenon is called the
"superconducting proximity effect". A similar question arises if one condenses
a thin ferromagnetic metal onto a normal metal. If the first layers of the
ferromagnet loose their magnetism they are denoted as "magnetic dead layers"
(MDL). The first dead layers were observed almost 30 years ago for two to
three atomic layers of Ni condensed onto amorphous Bi films \cite{B12}. For Ni
layers on top of noble metals there were originally contradicting results.
Liebermann et al. \cite{M86} observed two dead layers of Ni on Cu and Au while
Pierce and Siegmann \cite{P24} observed ferromagnetism already in mono-layers
of Ni on Cu (by means of spin polarized photo electrons). Kramer and Bergmann
\cite{B95} investigated the magnetic properties of Ni on the surface of Mg,
In, Sn, and the noble metals Cu, Ag and Au by means of the anomalous Hall
effect (AHE). They observed between two and three dead Ni layers on the
polyvalent substrate, while Ni on top of the noble metals showed
ferromagnetism already for the first Ni mono-layer. However, the electronic
properties of the Ni appeared to be modified for the first two to three Ni
layers because the AHE had the wrong (positive) sign in this range of Ni
thickness. Meservey et al. \cite{M90}, \cite{M87} used spin-polarized
tunneling to investigate the proximity effect of the ferromagnetic metals.
They observed about three dead layers of Ni on Al. A number of theoretical
papers \cite{T21}, \cite{F45}, \cite{F47}, \cite{Z21}, \cite{R31}, \cite{F46},
\cite{E16} investigated the question of a magnetic proximity effect.

The occurance of magnetic dead Ni layers is still a question under
investigation \cite{K71}, \cite{B147} while double layers of PbNi and other
pairs of superconductor and ferromagnetic metal experienced a new interest in
the superconducting proximity effect \cite{D25}. But in this paper we want to
address a claim made by Moodera and Meservey (MM) \cite{M88}, \cite{M85},
\cite{M84}, \cite{M89}, \cite{M83}, \cite{M91} about the properties of single
and double layers of Ni on Pb.

MM increased the sensitivity in their investigation of PbNi double layers by
using a $9nm$ thick Pb film as part of $14MHz$ oscillator. The frequency of
the oscillator changed by about $60kHz$ when the Pb film made a transition
from superconducting to normal state. They observed that the deposit of
$0.4nm$ of Ni onto the Pb substrate reduced the transition temperature of the
Pb below $4.2K$. A similar effect can be produced by the deposition of Fe.
However, the pair breaking effect of Fe is about $80$ times stronger than that
of Ni.

MM gave their results the following interpretation: The Ni atoms on top of the
Pb films do not completely loose their moments, even for the smallest
coverages, and their magnetic scattering dephases (depairs) the
superconducting Cooper pairs, even at their smallest thickness of $0.2$ atomic
layers of Ni. MM did not try to give a value for this reduced moment.

In this paper we revisit the PbNi system. We have measured the magnetic
scattering of Ni using the method of weak localization (quantum interference).
It is well known and discussed below that the pair-breaking mechanism in
superconductivity and the dephasing in weak localization are in many aspects
identical. There is, however, an experimental difficulty in measuring weak
localization in superconducting Pb films since the magnetoresistance is
overshadowed by the Azlamazov-Larkin fluctuations \cite{B32}. Therefore we use
only very thin Pb films between 1 and 10 atomic layers which are condensed
onto a Ag film. Then the proximity effect suppresses the superconductivity of
the Pb layers.

The experimental procedure is the following. A Ag film with a thickness of
$35$ atomic layers is quench condensed onto a quartz plate at He temperature
in an UHV of better than $10^{-11}$ torr. The Ag film is covered in different
experiments with Pb films whose thicknesses lie between one and five atomic
layers. Then the Pb film is covered in several steps with Ni. The first Ag
film is chosen for three reasons: (i) Even for quenched condensation it is not
possible to condense a homogeneous film of a few atomic layers of Pb onto a
quartz plate. This requires a homogeneous conducting metal film of sufficient
thickness as a substrate. (ii) The Ag film suppresses the superconductivity of
the extremely thin Pb film. (iii) Ni on top of Ag shows magnetism already for
a mono-layer of Ni. Therefore the observation of MDLs of Ni on AgPb can only
be due to the Pb film in between the Ag and the Ni. (It also proves that there
are no holes in the thin Pb film).

We use two experimental methods to investigate the magnetic properties of the
AgPbNi multi-layers, (i) the anomalous Hall effect and (ii) weak localization.
In Fig.1 the anomalous Hall resistance $R_{yx}\left(  B\right)  $ \ is plotted
as a function of $B$ for $d_{Pb}=2$ atomic layers and different Ni thicknesses
in a perpendicular magnetic field. The AHE curves can be extrapolated to zero
magnetic field, yielding $R_{yx}^{0}$. This AHE resistance $R_{yx}^{0}$ is
plotted in Fig.2. It measures the magnetization perpendicular to the film
plane in zero magnetic field. The surprising result is that the thickness of
the MDL is (almost) independent of the Pb thickness. Even for $1.3$ atomic
layers of Pb the AHE is suppressed up to a Ni thickness of $2.5$ \ atomic
layers The results are collected in table I.%

\[%
\begin{array}
[c]{cc}%
{\includegraphics[
height=2.4782in,
width=3.0154in
]%
{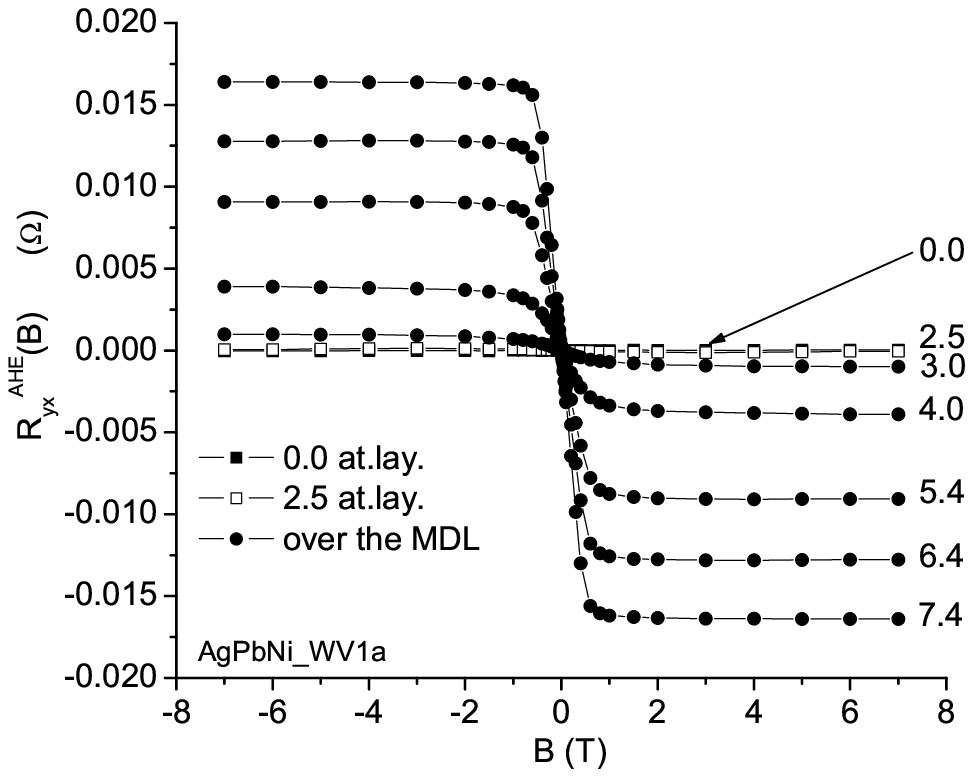}%
}%
&
\end{array}%
\begin{tabular}
[c]{l}%
Fig.1: The anomalous Hall curves\\
of an AgPbNi layer as a function of\\
the Ni thickness.
\end{tabular}
\]%
\[%
\begin{array}
[c]{cc}%
{\includegraphics[
height=3.6895in,
width=2.7015in
]%
{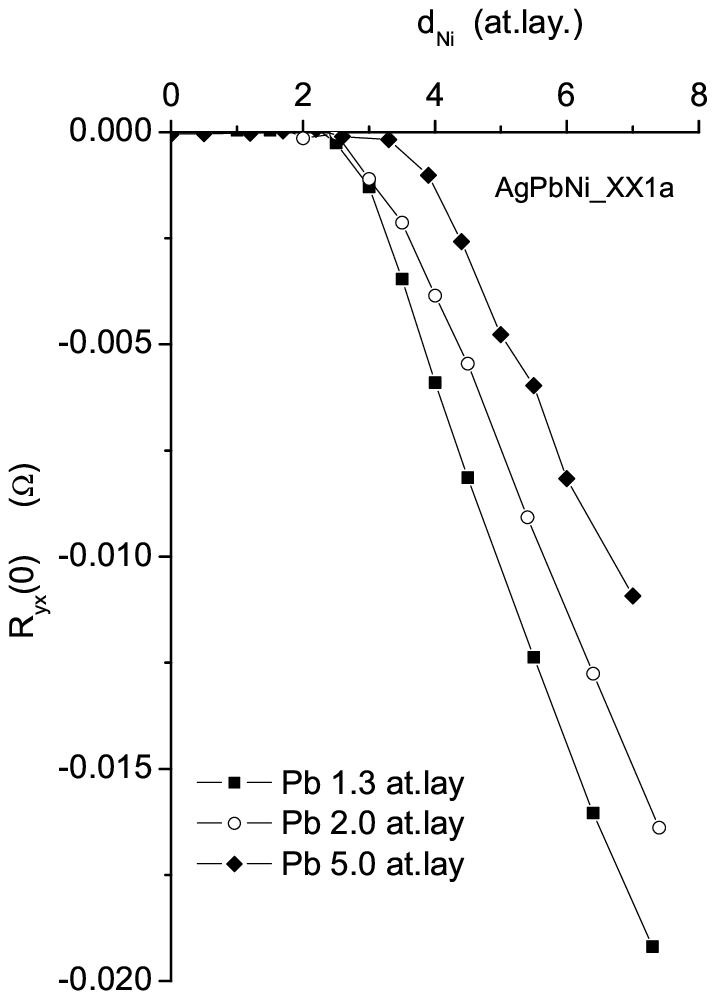}%
}%
&
\end{array}%
\begin{tabular}
[c]{l}%
Fig.2: The anomalous Hall resistance\\
$R_{yx}^{0}$ as a function of the $d_{Ni}$ for three\\
different Pb thicknesses.
\end{tabular}
\]%
\[%
\begin{array}
[c]{cc}%
\begin{tabular}
[c]{|l|l|}\hline
\textbf{d}$_{\text{\textbf{Pb}}}$ & \textbf{d}$_{\text{\textbf{MDL}}}$\\\hline
1.3 & 2.5\\\hline
2.0 & 2.5\\\hline
5.0 & 3.3\\\hline
\end{tabular}
&
\end{array}%
\begin{tabular}
[c]{l}%
Table I: The number of magnetic dead Ni layers\\
in AgPbNi multi-layers as measured by the\\
anomalous Hall effect. The first column gives\\
the thickness of the Pb film in atomic layers
\end{tabular}
\]

The experimental result confirms in addition that the Pb film covers the Ag
substrate homogeneously because one would otherwise have regions of magnetic
Ni on Ag which would be observable.

The magneto-resistance measurements of weak localization yield the dephasing
of the conduction electrons in the multi-layers. In Fig. 3 the
magneto-resistance of a AgPbNi multi-layer with a Pb thickness of $1.3$ atomic
layers is plotted for different Ni coverages. The numbers at the right side of
the curves gives the Ni coverage in atomic layers. The open and closed circles
are experimental data while the full curves are theoretical fits with two
fitting parameters, $H_{i}^{\ast}$ and $H_{so}^{\ast}$.%
\begin{align*}
&
{\includegraphics[
height=3.0079in,
width=3.6197in
]%
{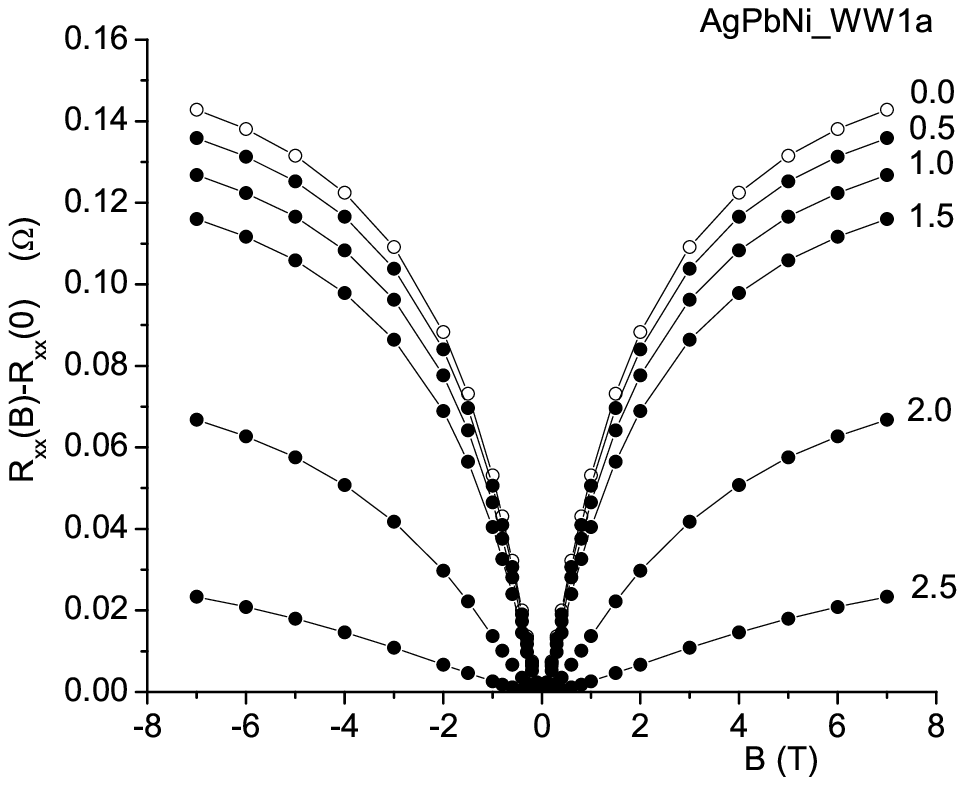}%
}%
\\
&
\begin{tabular}
[c]{l}%
Fig.3: Magneto-resistance of a AgPbNi multi-layer with\\
$d_{Pb}=1.3$ atomic layers The open and full circles are\\
experimental data while the full curves are theoretical fits,\\
yielding the singlet and triplet dephasing rates. The\\
numbers at the right of the curves give the Ni thickness in\\
atomic layers.
\end{tabular}
\end{align*}

The theory yields the following expression \cite{H1} for the quantum
conductance corrections to the resistance
\begin{equation}
\frac{\Delta R}{R^{2}}=\frac{e^{2}}{2\pi^{2}\hbar}\left[  \frac{1}{2}f\left(
\frac{B}{B_{S}}\right)  -\frac{3}{2}f\left(  \frac{B}{B_{T}}\right)  \right]
\label{dR}%
\end{equation}
where $R$ is the resistance per square and $\Delta R/R^{2}=-\Delta G$ is the
(negative) change in conductance. The function $f\left(  x\right)  $ is given
by
\[
f\left(  x\right)  =\ln\left(  x\right)  +\Psi\left(  \frac{1}{2}+\frac{1}%
{x}\right)
\]
where $\Psi\left(  z\right)  $ is the digamma function and $B_{S}$ and $B_{T}$
are the characteristic fields for singlet and triplet dephasing. They are
given by
\[%
\begin{array}
[c]{ccc}%
B_{S}=B_{i}+2B_{s}=B_{i}^{\ast} &  & B_{T}=B_{i}+\frac{4}{3}B_{so}+\frac{2}%
{3}B_{s}=B_{i}^{\ast}+\frac{4}{3}B_{so}^{\ast}%
\end{array}
\]
$B_{i},B_{so},B_{s}$ are the characteristic fields for inelastic, spin-orbit
and magnetic scattering. From these fields one can calculate the corresponding
scattering times $\tau_{n}$ with the relation
\[
B_{n}\tau_{n}=\frac{\hbar}{4eD}
\]
where $D$ is the diffusion constant which can be obtained from the resistivity
of the layers (see also \cite{L9}, \cite{B30}). This product has a value of
$B_{n}\tau_{n}$ $\thickapprox3\times10^{-13}Ts$ (it varies slightly for
different multi-layers).

In Fig.4 the (additional) dephasing field $1/\tau_{\varphi}$ due to the Ni
layers is plotted as a function of the Ni thickness $d_{Ni}$. The three plots
with full circles represent three different Pb layers in between the Ag and Ni
films (the numbers next to the curves give the Pb thickness in atomic layers).
For comparison the additional dephasing for Ni directly deposited on Ag is
shown (full squares). Here the Ni possesses a magnetic moment already in the
first mono-layer.%

\begin{align*}
&
\begin{array}
[c]{cc}%
{\includegraphics[
height=3.1532in,
width=3.8049in
]%
{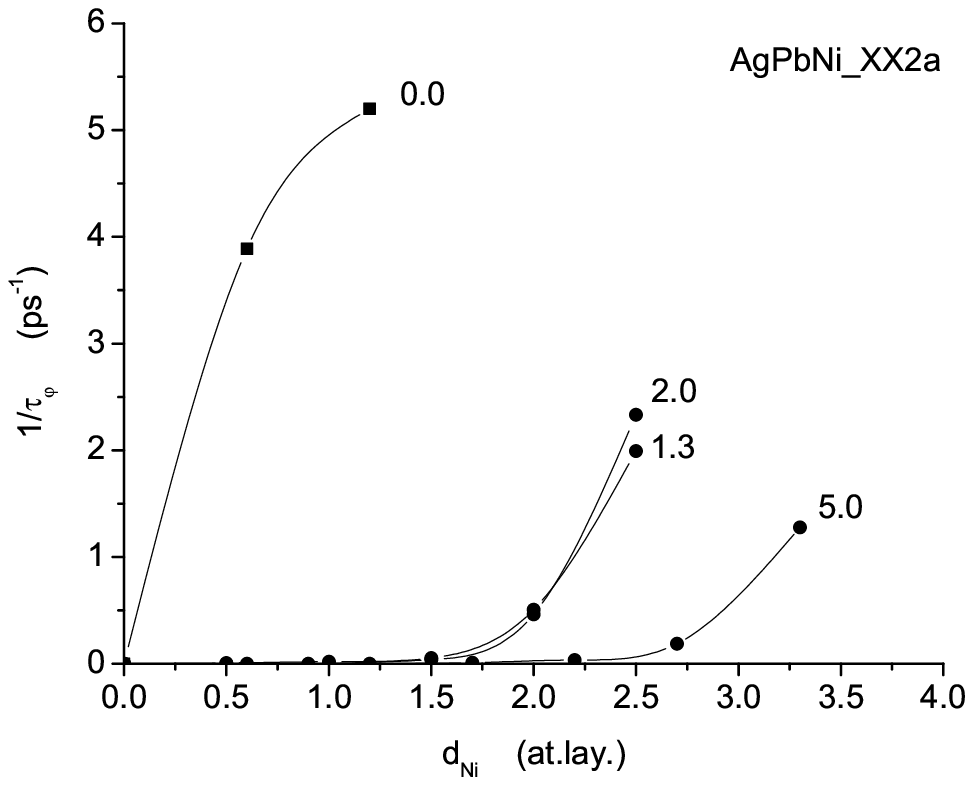}%
}%
&
\end{array}%
\begin{tabular}
[c]{l}%
Fig.4: The additional dephasing\\
rate $1/\tau_{\varphi}$ of AgPbNi multi-layers\\
as a function of $d_{Ni}$ for different\\
Pb thicknesses $d_{Pb}$ (full circles).\\
The full squares give $B_{s}$ for AgNi\\
layers.
\end{tabular}
\\
&
\end{align*}

From Fig.3 and Fig.4 it is very obvious that the additional dephasing of Ni on
Pb is very small for Ni thicknesses up to $1.5$ atomic layers The dephasing
rate for the AgPbNi multi-layer with $d_{pb}=2.0$ atomic layers and
$d_{Ni}=1.5$ atomic layers is $1/\tau_{\varphi}\thickapprox3.\,\allowbreak
0\times10^{10}s^{-1}.$ This is the average dephasing rate for an electron that
propagates within the multi-layer AgPbNi. Since the electron spends only the
fraction $d_{Ni}N_{Ni}/\left(  d_{Ag}N_{Ag}+d_{Pb}N_{Pb}+d_{Ni}N_{Ni}\right)
$ of the time in the Ni films the actual dephasing in the Ni film is larger by
the factor $\left(  d_{Ag}N_{Ag}+d_{Pb}N_{Pb}+d_{Ni}N_{Ni}\right)
/d_{Ni}N_{Ni}\thickapprox d_{Ag}N_{Ag}/d_{Ni}N_{Ni}$. This yields for the
dephasing in the Ni $1/\tau_{Ni}^{wl}=3.\,\allowbreak0\times10^{10}s^{-1}\ast
d_{Ag}N_{Ag}/d_{Ni}N_{Ni}$.

Moodera and Meservey interpreted the $T_{c}$-reduction in their PbNi double
layers to be a result of magnetic scattering in the Ni film which causes a
dephasing or depairing of the Cooper pairs. MM observed that a coverage with
$1.5$ atomic layers of Ni yielded a reduction of $T_{c}$ in their Pb film of
about $\Delta T_{c}\thickapprox2.5K$. This $T_{c}$-reduction is supposed to be
due to a (magnetic) dephasing rate $\tau_{\varphi}$. For a weak couling
superconductor a dephasing rate of $1/\tau_{\varphi}$ yields (in linear
approximation) a $T_{c}$-reduction of $k_{B}\Delta T_{c}=\pi\hbar/\left(
8\tau_{\varphi}\right)  $. For the strong coupling superconductor Pb one
obtains a correction factor of $1.4$ \cite{B44} which is neglected here. This
yields a dephasing rate of $1/\tau_{\varphi}\thickapprox8k_{B}\Delta
T_{c}/\left(  \hbar\pi\right)  \thickapprox\allowbreak8.\times10^{11}s^{-1}$
in the double layer PbNi with $d_{Ni}=1.5$ atomic layers As before the
corresponding dephasing in the Ni film is larger by the factor $d_{Pb}%
N_{Pb}/d_{Ni}N_{Ni}$, $1/\tau_{Ni}^{SU}=\allowbreak8.\times10^{11}s^{-1}%
d_{Pb}N_{Pb}/d_{Ni}N_{Ni}$.

Next we compare the dephasing rates in the Ni between the superconducting and
the weak localization measurement. The ratio of the two rates is%

\[
r=\frac{\left(  1/\tau_{Ni}^{SU}\right)  }{\left(  1/\tau_{Ni}^{wl}\right)
}=\frac{8\times10^{11}s^{-1}\ast d_{Pb}^{MM}\gamma_{Pb}}{3.2\times
10^{10}s^{-1}\ast d_{Ag}^{TB}\gamma_{Ag}}=109
\]

Here we have replaced the ratio $N_{Pb}/N_{Ag}$ by the ratio of their
Sommerfeld constants $\gamma_{Pb}/\gamma_{Ag}=4.\,\allowbreak4$. $d_{Pb}%
^{MM}=9nm$ is the Pb thickness in MM's experiment and $d_{Ag}^{TB}=9nm$ is the
the Ag thickness in our experiment. Obviously the dephasing rate calculated
from the reduction of $T_{c}$ in Pb is much larger than the observed magnetic
dephasing in weak localization.

To better understand this great difference in the dephasing rates we compare
the "pair propagator" in superconductivity and weak localization. Fig.5a shows
the pair propagator of a Cooper pair in superconductivity. One electron with
spin up moves from A to B and its time reversed partner with spin down travels
from B to A. Fig.5b shows the closed loop along which two partial waves of a
single electron travel in opposite directions. The fact that the path in weak
localization is closed is not important for the dephasing processes. What is
important is that both propagators move along time reversed paths.

Let us start with weak localization. Here the clockwise and the counter-clock
wise moving electron waves interact with the environment. Electron-phonon
processes, Coulomb interaction with the other electrons, and magnetic
scattering cause changes in energy and phase between the two partial waves.
However, there is no interaction between the two partial waves because they
belong to the same electron.

In the superconducting case we have actually two electrons. Each one
experiences the same interaction with the environment as in weak localization
and suffers the same dephasing. For these processes the dephasing of weak
localization is the same as the pair breaking in superconductivity.

But in superconductivity there are additional processes which influence the
phase coherence since there are two electrons which interact. This influences
their coherence in several ways: (i) the exchange of phonons supports the
coherence and causes the superconductivity,(ii) due to the finite temperature
the coherence of the two electrons decays with time as $%
{\textstyle\sum_{n}}
\exp\left[  -2\left\vert \omega_{n}\right\vert t\right]  $ ($\omega_{n}=2\pi
k_{B}T\left(  n+1/2\right)  /\hbar$ =Matsubara frequencies). Since the balance
of (i) and (ii) determines the $T_{c}$ of the superconductor these two
processes are already taken care of.

But there can be Coulomb interaction between the two electrons which causes
dephasing or pair weakening. This effect is not present in weak localization.
The pair propagator (or pair amplitude) propagates also into the Ni film. If
the two electrons of the pair hop into a d-state of a Ni atom and overlap,
they repel each other and weaken the superconductivity.

If one adds transition metal atoms to a superconductor, either as impurities
or as a thin film then one seriously alters the properties of the
superconductor. The d-atoms possess d-resonance states. One can divide them
into three groups : (i) no Coulomb interaction between d-electrons of opposite
spin, (ii) a finite Coulomb interaction between d-electrons of opposite spin,
but no magnetic moment, (iii) the Coulomb interaction is sufficiently strong
to break the symmetry between opposite spins, and a magnetic moment develops.
It is well known the third case causes pair breaking. In addition, it has been
shown that not only case (ii) but also case (i) \ reduces the transition
temperature. This phenomenon has been studied in 1970's (see for example
\cite{Z24}, \cite{B128}, \cite{S40}) and is often called pair weakening.

It might be surprising that one obtains a $T_{c}$-reduction even in the
absence of Coulomb repulsion, ut its physical origin is quite simple. When a
Cooper pair jumps onto two (time reversed) d-states of an d-atom their pair
amplitude still decays due to the finite temperature, but it is screened from
the attractive electron-phonon interaction. The fraction of the time on the
d-resonances is roughly $N_{d}/N_{fe}$ where $N_{d}$ is the density of
states\ of the d-resonance of all d-atoms while $N_{fe}$ is the density of
states of the (free) conduction electrons. As a consequence the attractive
interaction is reduced by a factor $\left(  1-N_{d}/N_{fe}\right)  $. This
yields a reduction in $T_{c}$.%

\begin{align*}
&
{\includegraphics[
height=1.6812in,
width=4.7206in
]%
{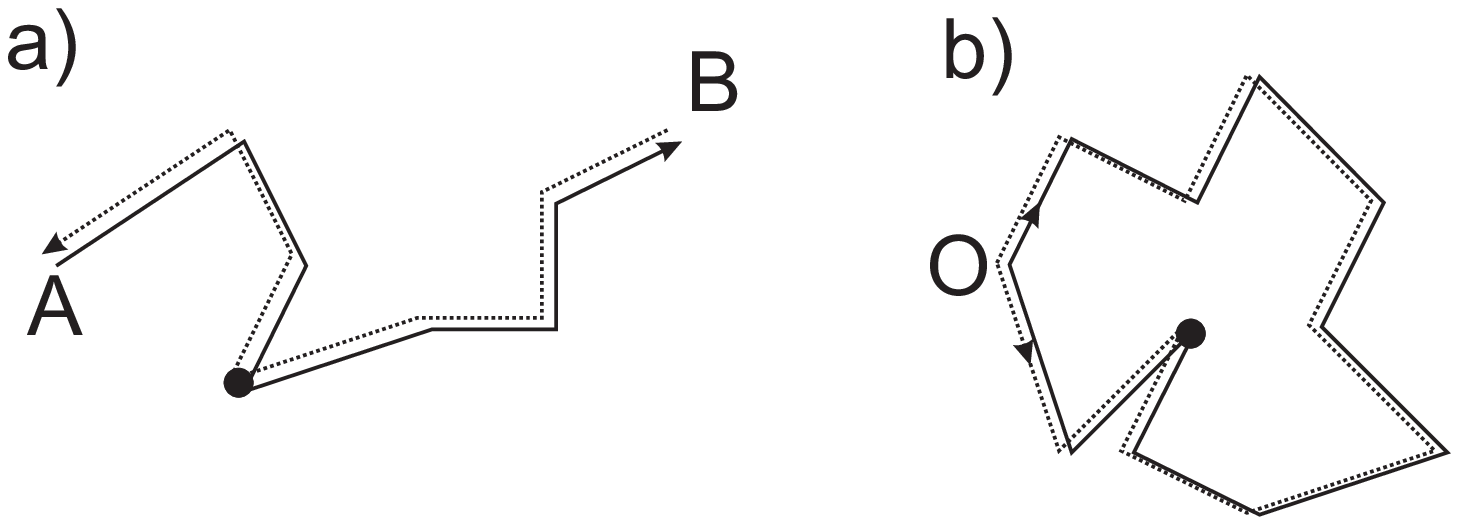}%
}%
\\
&
\begin{tabular}
[c]{l}%
Fig.5a: The pair propagator of a Cooper pair in a superconductor.\\
Fig.5b: The particle-particle propagator in weak localization.
\end{tabular}
\end{align*}

We conclude that our experimental results and those by Moodera and Meservey do
not contradict each other. However, our experimental results contradict their
conclusion. We find a much smaller dephasing than the reduction of $T_{c}$
would suggest. We conclude that reduction of the superconducting transition
temperature is not due to dephasing by magnetic scattering but due to the
resonance scattering of Cooper pairs by non-magnetic d-states. A Coulomb
repulsion as studied by Friedel and Anderson will enhance the $T_{c}$-reduction.

Actually the combined investigation of magnetic dead Ni layers on top of a
superconductor, the reduction of $T_{c},$ and the dephasing of weak
localization would be a very effective method to study the formation of
magnetic moments in d-metals.

Acknowledgement: The research was supported by the National Science Foundation
DMR-0439810.


\newpage

\begin{thebibliography}{99}                                                                                               %
\bibitem {B12}G. Bergmann, Phys. Rev. Lett. 41, 264 (1978)\newline

\bibitem {M86}L. Liebermann, J. Llinton, D. M. Edwards, and J. Mathon, Phys.
Rev. Lett. 25, 232 (1970) \newline

\bibitem {P24}D. T. Pierce and H. C. Siegmann, Phys. Rev. B9, 4035
(1974)\newline

\bibitem {B95}I. Kramer, and G. Bergmann, Phys. Rev. B27, 7271 (1983)\newline

\bibitem {M90}R. Meservey, P. M. Tedrow, and V. R. Kalvey, Solid state Comm.
36, 969 (1980) \newline

\bibitem {M87}R. Meservey, P. M. Tedrow, and V. R. Kalvey, J. Appl. Phys. 52,
1617 (1981) \newline

\bibitem {T21}B. N. Cox, R. A. Tahir-Kheli, and R. J. Elliot Phys. Rev. B 20,
2864 (1979) \newline

\bibitem {F45}D. S. Wang, A. J. Freeman, and H. Krakauer, Phys. Rev. B 24,
1126 (1981) \newline

\bibitem {F47}J. Tersoff and L. M. Falicov, Phys. Rev. B 26, 6186 (1982)
\newline

\bibitem {Z21}C. Uher, R. Clarke, G. Zheng, and I. K. Schuller, Phys. Rev. B
30, 453 (1984) \newline

\bibitem {R31}R. Richter, J. G. Gay, and J. R. Smith, Phys. Rev. Lett. 54,
2704 (1985) \newline

\bibitem {F46}S. C. Hong, A. J. Freeman, and C. L. Fu, Phys. Rev. B 39, 5719
(1989) \newline

\bibitem {E16}A. Ernst, M. Lueders, W. M. Temmerman, Z. Szotek, and G. V Laan,
J. Phys. Condens. Matter 12, 5599 (2000) \newline

\bibitem {K71}S. Kim, J. Jeong, J. B. Kortright, and S. Shin, Phys. Rev. B 64,
052406 (2001) \newline

\bibitem {B147}P. SanGiorgio, S. Reymond, M. R. Beasley, J. H. Kwon, and K.
Char, Phys. Rev. Lett. 100, 237002 (2008) \newline

\bibitem {D25}O. Bourgeois, R. C. Dynes, Phys. Rev. B65, 144503 (2002)
\newline

\bibitem {M88}J. S. Moodera, and R. Meservey, Phys. Rev. B 29, 2943 (1984)
\newline

\bibitem {M85}J. S. Moodera, R. Meservey, and P. M. Tedrow, J. Appl. Phys. 55,
2502 (1984) \newline

\bibitem {M84}J. S. Moodera and R. Meservey, Phys. Rev. Lett. 55, 2082 (1985)
\newline

\bibitem {M89}J. S. Moodera, and R. Meservey, Solid state Comm. 29, 295 (1986)
\newline

\bibitem {M83}J. S. Moodera and R. Meservey, Phys. Rev. B 34, 379 (1986)
\newline

\bibitem {M91}J. S. Moodera, and R. Meservey J. Appl. Phys. 61, 3741 (1987)
\newline

\bibitem {B32}G. Bergmann, Phys. Rev. B29, 6114 (1984)\newline

\bibitem {H1}S. Hikami, A. I. Larkin and Y. Nagaoka, Prog. Theor. Phys. , 63,
707 (1980) \newline

\bibitem {L9}P. A. Lee and T. V. Ramakrishnan, Rev. Mod. Phys. 57, 287
(1985)\newline

\bibitem {B30}G. Bergmann, Physics Reports 107, 1 (1984)\newline

\bibitem {B44}D. Rainer, and G. Bergmann, J. Low Temp. Phys. 14, 501
(1974)\newline

\bibitem {Z24}M. J. Zuckermann, Phys. Rev. 140, A899 (1965) \newline

\bibitem {B128}C. F. Ratto, and A. Blandin, Phys. Rev. 156, 513 (1967)
\newline

H. Shiba, Prog. Theor. Phys. 50, 50 (1973) \newline
\end{thebibliography}
\end{document}